\documentclass[twocolumn]{aastex702}

\usepackage[dvipsnames]{xcolor}
\usepackage{amsmath}
\usepackage{bm}

\begin{document}

\title{Initial Conditions Dictate Gas-Stellar Counter-Rotation in Galaxies}

\author[gname=Lin, sname='Zhu']{Lin Zhu}
\affiliation{Department of Astronomy, Xiamen University, Xiamen, Fujian 361005, China}
\email{linzhu21@stu.xmu.edu.cn}  

\author[gname=Hao-Ran, sname='Yu']{Hao-Ran Yu} 
\affiliation{Department of Astronomy, Xiamen University, Xiamen, Fujian 361005, China}
\email[show]{haoran@xmu.edu.cn}

\author[gname=Min, sname=Bao]{Min Bao}
\affiliation{School of Physics and Technology, Nanjing Normal University, Nanjing, 210046, China}
\email{mbao@nnu.edu.cn}

\author[gname=Ming-Jie, sname=Sheng]{Ming-Jie Sheng}
\affiliation{Department of Astronomy, Xiamen University, Xiamen, Fujian 361005, China}
\email{mingjie@xmu.edu.cn}

\author[gname=Jian, sname=Qin]{Jian Qin}
\affiliation{Department of Astronomy, Xiamen University, Xiamen, Fujian 361005, China}
\affiliation{Center for Astronomy and Space Sciences, China Three Gorges University, Yichang 443000, China }
\affiliation{College of Mathematics and Physics, China Three Gorges University, Yichang 443000, China}
\email{qinjian@ctgu.edu.cn}

\author[gname=Shuai, sname=Lu]{Shuai Lu}
\affiliation{Department of Astronomy, Xiamen University, Xiamen, Fujian 361005, China}
\email{lushuai@stu.xmu.edu.cn}

\author[gname=Min, sname=Du]{Min Du}
\affiliation{Department of Astronomy, Xiamen University, Xiamen, Fujian 361005, China}
\email{}

\author[gname=Yanmei, sname=Chen]{Yanmei Chen}
\affiliation{School of Astronomy and Space Science, Nanjing University, Nanjing 210023, China}
\affiliation{Key Laboratory of Modern Astronomy and Astrophysics (Nanjing University), Ministry of Education, Nanjing 210023, China}
\email{chenym@nju.edu.cn}

\author[gname=Yipeng, sname=Jing]{Yipeng Jing}
\affiliation{State Key Laboratory of Dark Matter Physics, Tsung-Dao Lee Institute $\&$ School of Physics and Astronomy,  Shanghai Jiao Tong University, Shanghai 201210, China}
\email{}

\begin{abstract}
We present a physical origin of gas-stellar counter-rotation in galaxies.
Scale-dependent partitions of tidal torques from cosmic primordial perturbations source angular momentum for distinct baryonic components. This scale dependence maps to radial hierarchy in Lagrangian space clustering: stellar components, assembled at earlier times from the inner protohalo, retain persistent memory of small-scale tidal torques, whereas gas, accreted later from outer Lagrangian regions, is governed by large-scale perturbations and can develop counter-rotation. This mechanism is observationally testable and offers a direct probe of the hierarchical cold dark matter (CDM) paradigm. Unlike conventional late-time scenarios that only explain the existence of counter-rotation, our framework further predicts its component-resolved direction, enabling component-resolved angular momentum to serve as a new probe of primordial density perturbations and improving cosmological initial condition reconstruction.
\end{abstract}


\section{Introduction} 
Cosmological information from the large-scale structure of the universe is a fundamental goal of modern astrophysics. Beyond spatial distribution and peculiar velocities, galaxy angular momenta (hereafter spins) emerge as an independent probe of the cosmological initial conditions. Tidal torque theory provides the standard framework for the origin of galaxy spins, both in direction and magnitude \citep{Peebles1969ApJ,Doroshkevich1970Afz,White1984ApJ}, and has been verified in numerical simulations \citep{Porciani2002MNRAS} and observations. The latter additionally relies on the recovery of the cosmic initial conditions by using reconstruction techniques \citep{Wang2014ApJ}, and a field-level spin estimator, which directly derives the spin observable from primordial perturbations in Lagrangian space \citep{Yu2020PhRvL}. Using the ELUCID project \citep{2016WangApJ} reconstructing the initial conditions of the local universe from the SDSS galaxy distribution \citep{York2000AJ}, and the spin estimator \citep{Yu2020PhRvL} as an approximation of the tidal torque theory, the spin (direction) conservation of initial tidal torques all the way to different galaxy types and wide range of host halo masses has been verified \citep{Motloch2021NA,2026ShengNA}. 

The above detection relies on the fact that baryons trace dark matter (DM) spins. In general, diverse baryonic matter (e.g. gas, stars, and on different scales) co-rotate with dark matter \citep{Sheng2023ApJ}. Meanwhile, a small portion of gas–star counter-rotating galaxies (CRGs) are detected both in simulations \citep{Osman2017MNRAS,Khoperskov2021MNRAS,C-V2026arXiv260607160C} and in observations \citep{Chen2016NatCo,Jin2016MNRAS,Zhou2022MNRAS,Xu2022MNRAS,zhang2025ApJ,Gasymov2025ApJS}. Their origin is usually attributed to stochastic late-time processes, such as retrograde gas accretion or mergers \citep{Starkenburg2019ApJ,Zhou2022MNRAS,Zinchenko2023A&A,Beom2024AJ,Bao2025ApJL,Bugue2026A&A}. However, these late-time scenarios only explain the existence of counter-rotation, leaving both its component-resolved directional specifics and its potential as a cosmological probe largely unexplored. 

In this Letter, we trace the multi-component spin evolution of central galaxies back to their Lagrangian initial conditions and demonstrate that gas-stellar counter-rotation is fundamentally determined by the scale dependence of primordial tidal torques.

\section{Methods} \label{sec:methods} 
The spin vector $\bm{J}$ of any given set of matter $M$ is given by $(\bm{J}^M)_i\equiv\int_M\epsilon_{ijk}\eta_j \dot\eta_k{\rm d}M$, where $\epsilon_{ijk}$ is the 3D Levi-Civita symbol, $\eta_i=(\bm{\eta})_i$ is the comoving coordinate of the mass element ${\rm d}M$ relative to $M$'s center of mass (CoM), and overdot denotes the time derivative. Specifically, in Lagrangian space, the coordinates are just the ones when scale factor $a$ approaches zero, $\eta_i=\eta_i(a\to 0)$, and Lagrangian spin $\bm{J}^M_{\rm L}=\bm{J}^M_{\rm E}(a\to 0)$, where $\bm{J}^M_{\rm E}(a)$ is the Eulerian spin of $M$ at $a$. 

Based on the tidal torque theory and the field-level spin estimator \citep{Yu2020PhRvL},\footnote{We use $\bm J$ for spin of an object, and $\bm L$ for a spin field.}
\begin{eqnarray}
\left(\bm{J}_{\rm IT}^M\right)_i &=& \epsilon_{ijk} I_{jl}^M T_{lk}^M,\label{eq.IT}\\
\left(\bm{L}_{\rm TT}\right)_i &=& \epsilon_{ijk} T_{jl}(\bm{q},R)T_{lk}(\bm{q},R_+),\label{eq.TT}
\end{eqnarray}
where $\mathbf{I}^M=(I_{ij}^M)\equiv\int_M\eta_i\eta_j{\rm d}M$ is the inertia tensor of $M$, $\mathbf{T}^M=(T_{ij}^M)\equiv-\int_M\phi_{,ij}{\rm d}M$ is the tidal tensor $M$ feels,\footnote{Since only dimensionless spin magnitude is used in this work, we absorb unnecessary coefficients such as Newton's constant.} and $T_{ij}(\bm{q},R)$ is the tidal tensor at Lagrangian position $\bm{q}$, smoothed by scale $R$. 
We note that both $\bm{J}_{\rm IT}^M$ and $\bm{L}_{\rm TT}$ are derived in Lagrangian space, however $\bm{J}_{\rm IT}^M$ is a quantity of an object $M$, but $\bm{L}_{\rm TT}$ is a vector field in Lagrangian space, and only the latter can be derived directly from primordial perturbations. \citet{Yu2020PhRvL} find that if $R_+\to R$ and $R$ takes the Lagrangian radius of $M$, i.e., $R=\left(2MG/\Omega_{\rm m}H_0^2\right)^{1/3}$, $\bm{L}_{\rm TT}$ well approximates $\bm{J}_{\rm IT}^M$. 
In order to reduce the statistical uncertainty of poorly defined spin directions in low-spin systems, dimensionless spin parameters are introduced \citep{Wu2021PhRvD,Sheng2024PhRvD} to parameterize the rotational support in Eulerian ($\lambda_K^{\rm{E}}$) and Lagrangian ($\lambda_K^{\rm{L}}$) spaces as
\begin{eqnarray}
\lambda_K &\equiv& \frac{\int_M\hat{J}_i \epsilon_{ijk} \eta_j \dot{\eta}'_k \, \mathrm{d}M}{\int_M\eta \dot{\eta}' \, \mathrm{d}M} \in[0,1], \label{eq.LambdaK}
\end{eqnarray}
where $\dot\eta_i'\equiv\dot\eta_i-\bar{\dot\eta}_i$ is the velocity of ${\rm d}M$ relative to the mean velocity of $M$, $\eta\equiv|\bm{\eta}|$, ${\dot\eta}'\equiv|{\dot{\bm{\eta}}}'|$ and $\hat{J}_i\equiv J_i/|\bm{J}|$.  
The dimensionless intensity of the tidal torque is parametrized by
\begin{eqnarray}
\beta_{\rm IT} &=& \frac{|\epsilon_{ijk}I_{jl}T_{lk}|}{-I_{ij}T_{ij}},\label{eq.betaIT}\\
\beta_{\rm TT}(R) &=& \frac{|\epsilon_{ijk}\phi_{,jl}(R)\delta_{,lk}(R)|}{-\phi_{,ij}(R)\delta_{,ij}(R)},\label{eq.betaTT}
\end{eqnarray}
where $\delta$ is the primordial overdensity field. We use $\mu_{\alpha\beta}\equiv\hat{\bm{J}}_\alpha \cdot \hat{\bm{J}}_\beta\in[-1,1]$ to parameterize the directional correlation between vectors.

\section{sample selection} \label{sec:data}

We use the TNG100-1 simulation of the IllustrisTNG project \citep{Springel2018MNRAS, Pillepich2018MNRAS, Nelson2019ComAC}, which follows the coupled evolution of DM, gas, magnetic fields, and self-consistently formed stars and supermassive black holes within a periodic cosmological box of $75\,{\rm Mpc}\, h^{-1}$ per side to redshift $z=0$ ($a=1$). We restrict our analysis to a sample of $4069$ well-resolved central galaxies with total mass $M > 10^{11.5}\,M_\odot \, h^{-1}$ and stellar mass $M_* > 10^{9}\, M_\odot \, h^{-1}$. 

\begin{figure}[ht!]
\centering
\includegraphics[width=\columnwidth]{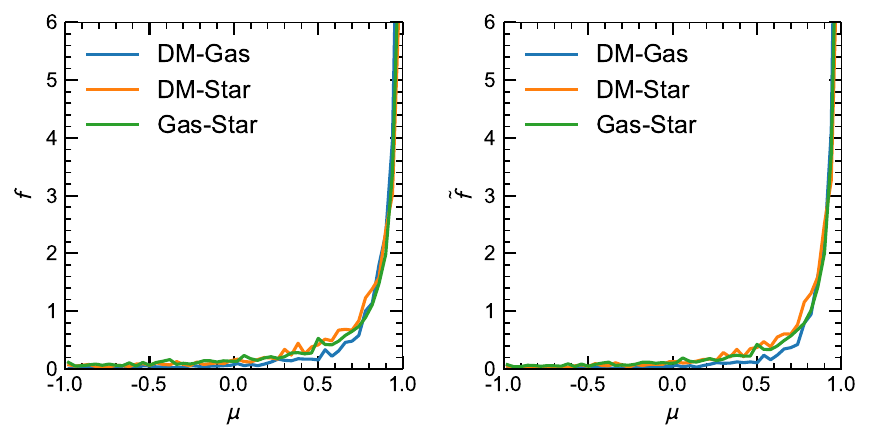}
\caption{Probability density functions (PDFs) of spin correlation between different galactic components. The two panels show the PDF ($f$) and the spin-parameter-weighted PDF ($\tilde{f}$) between DM, gas, stellar components respectively.}
\label{fig:pdf}
\end{figure}

\begin{figure*}[ht!]
\centering
\includegraphics[width=\textwidth]{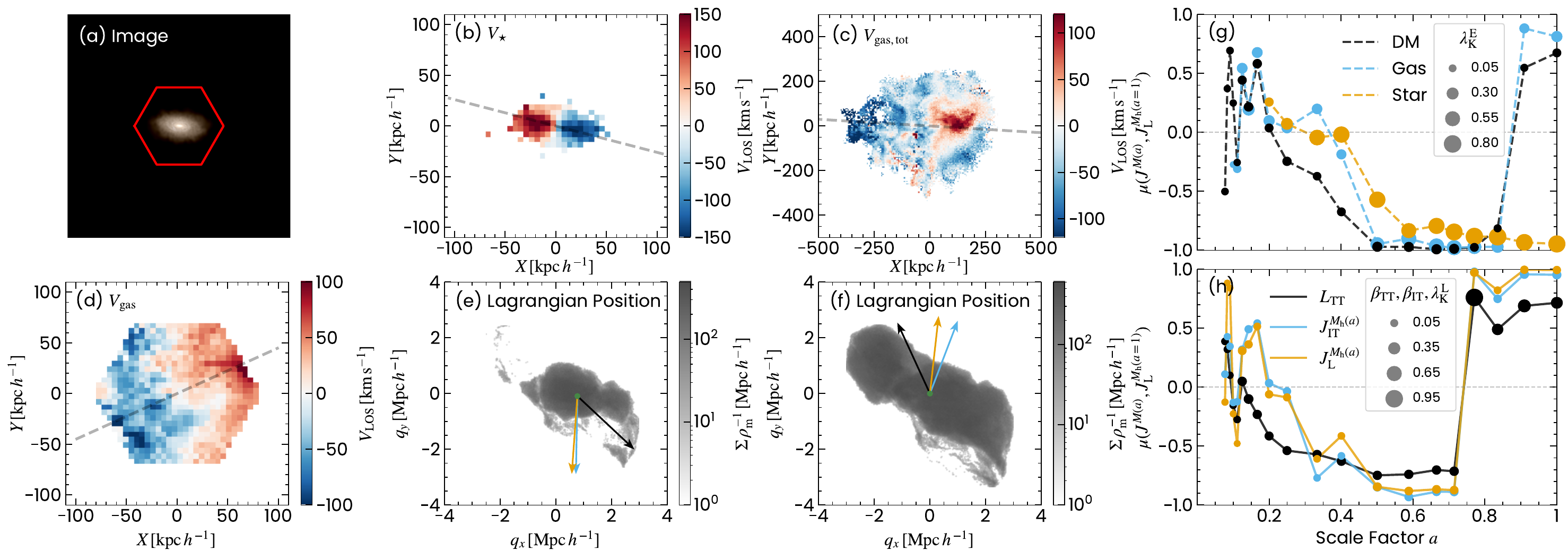}
\caption{Position, velocity and spin evolution of a representative CRG, subhaloID {\tt 393336}. Panel (a): mock SDSS g,\,r,\,z image. Panels (b\,-\,d): velocity maps of the stellar component within the field of view (FoV), total gas component, and gas component within FoV, where the dashed lines indicate their position angles. Panels (e, f): collapsed DM at $a\simeq 0.71$ and $a=1$, traced back to Lagrangian space, where the green dots mark the CoM, and the arrows denote the spin directions of the ${\bm L}_{\rm TT}({\bm q}(a),R(a))$ (black), ${\bm J}^{M_{\rm h}(a)}_{\rm IT}$ (blue) and ${\bm J}^{M_{\rm h}(a)}_{\rm L}$ (orange). Panels (g, h): component-wise spin correlation w.r.t. the initial spin direction of the whole system. The dot sizes indicate the intensity of the spin / tidal-torque, where spins with higher intensities are more reliable. 
}\label{fig.2}
\end{figure*}

Figure~\ref{fig:pdf} presents the distributions of the spin correlations between DM, gas, and stellar components. For sample selection, we construct a spin-parameter-weighted PDF in which each pair of distinct galactic components, $\alpha, \beta \in \{\mathrm{DM}, \mathrm{gas}, \mathrm{stars}\}$ with $\alpha \ne \beta$, is weighted by $\sqrt{\lambda_{K,\alpha}^{\rm E} \lambda_{K,\beta}^{\rm E}}$, thus down-weighting slow rotators whose spin directions are poorly defined. After discarding the lower $5\%$ tail of the $\lambda_K^{\rm E}$ distribution for each component, we eventually select 133 CRGs with $\mu_{\rm gas,star}<-0.5$. Among them, 108 show gas-DM co-rotation ($\mu_{\rm DM, gas}>0$). For these 133 CRGs, $\langle \mu_{\rm DM,gas}\rangle \simeq 0.54$ and $\langle \mu_{\rm  DM,star}\rangle \simeq -0.3$, indicating that the gas component typically shares a common kinematic origin with the DM halo, while the stellar spin is kinematically decoupled. Thus, we focus on the dominant CRGs, in which stars counter-rotate against both gas and DM.

To illustrate in detail, we select a representative counter-rotating galaxy (subhaloID {\tt 393336} in TNG100-1) with a stellar mass of $M_\star \simeq 4\times 10^{10}\,M_\odot \, h^{-1}$, comparable to the typical stellar mass of MaNGA galaxies $(M_\star \simeq 2 \times 10^{10}\,M_\odot \, h^{-1})$, and exhibiting a pronounced gas-stellar counter-rotation with $\mu_{\rm gas,star} \simeq -0.9$. We also select a typical co-rotating galaxy (subhaloID {\tt 480712} in TNG100-1) with a similar stellar mass $(M_\star \simeq 2 \times 10^{10}\,M_\odot \, h^{-1})$ for comparison.

\section{Results} \label{sec:results}
The mock SDSS image of subhaloID {\tt 393336} is displayed in Figure \ref{fig.2}(a), viewed within a reference frame where the gas spin aligns with the $y$-axis. The velocity maps (Panels b\,-\,d) confirm the opposite spins of the stellar and gas components. The velocity map of the entire gas component (Panel c) reveals a much more pronounced counter-rotation. We employ a time-dependent tracing method to investigate the spin origin of a given halo-galaxy system. Specifically, given epoch $a$, we identify its {\it collapsed progenitor} $M_{\rm h}(a)$ by the standard Friends-of-Friends algorithm, with linking parameter $b=0.2$, in Eulerian space, and then trace $M_{\rm h}(a)$ back to Lagrangian space. Following Section \ref{sec:methods}, the Lagrangian spin of $M_{\rm h}(a)$, its tidal torque and field-level spins are denoted as ${\bm J}^{M_{\rm h}(a)}_{\rm L}$, ${\bm J}^{M_{\rm h}(a)}_{\rm IT}$ and ${\bm L}_{\rm TT}({\bm q}(a),R(a))$.

For example, for {\tt 393336}, the two typical epochs are $a\simeq 0.71$ (peak star formation time) and $a=1$, and their corresponding collapsed Lagrangian patches are plotted in Panels (e,\,f). At these two epochs, the halo mass grew to $M_{\rm h}(a\simeq0.71)\simeq 6.65 \times  10^{11} M_\odot \, h^{-1}$ and $M_{\rm h}(a=1)\simeq 2.75 \times  10^{12} M_\odot \, h^{-1}$. Furthermore, since the Lagrangian patches remain largely compact and simply connected, their spatial extents can be naturally quantified by equivalent Lagrangian radii, $R(a\simeq0.71)\simeq 1.6\,{\rm Mpc}\,h^{-1}$ and $R(a=1)\simeq 2.6\,{\rm Mpc}\,h^{-1}$. Relative to the Lagrangian CoM of the halo at $a=1$, the CoM of the halo traced back from $a\simeq0.71$ reveals a distinct spatial displacement. Despite this, Panels (e,\,f) remarkably reveal a radially stratified structure, wherein the inner-to-outer Lagrangian regions inherently track the early-to-late formation epochs.
This spatial stratification is dictated by the hierarchical nature of cold DM.

\begin{figure*}[ht!]
\centering
\includegraphics[width=\textwidth]{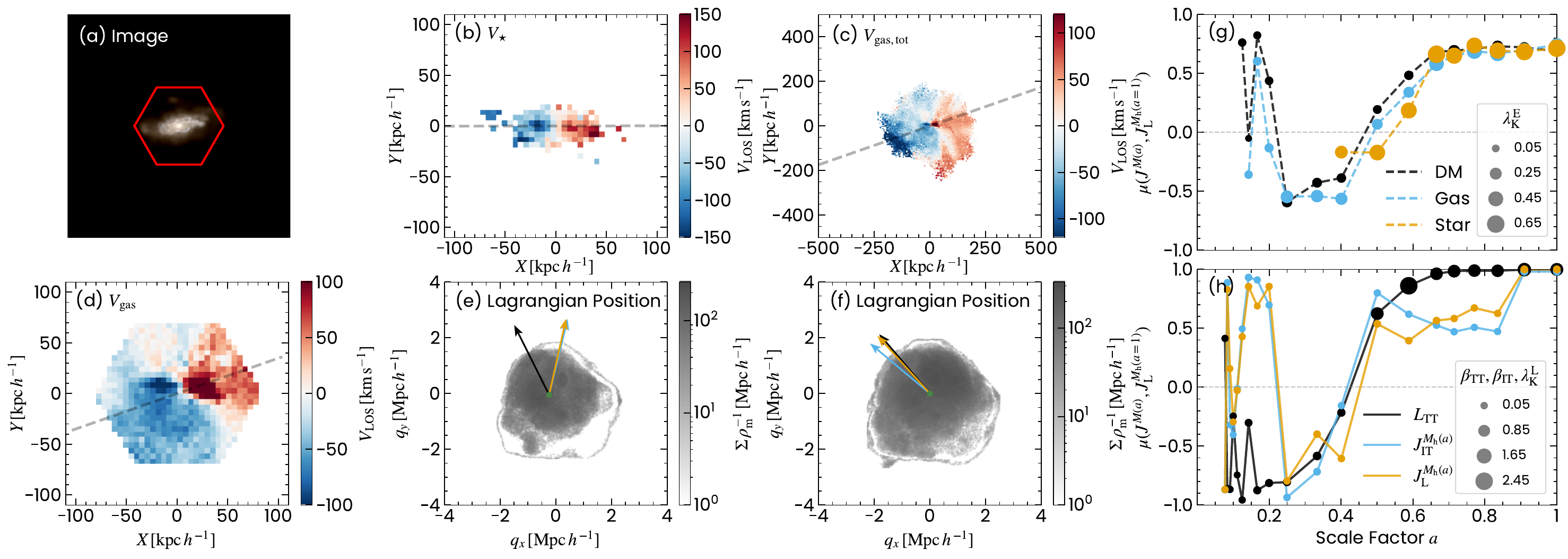}
\caption{Same as Figure 2, but for a representative co-rotating galaxy, subhaloID {\tt 480712}.}
\label{fig:3}
\end{figure*}

Using $\mathbf{I}^M$ and $\mathbf{T}^M$ extracted from the Lagrangian patches (Panels e,\,f), we derive the spins via Eqs. (\ref{eq.IT})--(\ref{eq.TT}) and find a remarkable consistency among these Lagrangian spins, such that ${\bm J}^{M_{\rm h}(a)}_{\rm L} \simeq {\bm J}^{M_{\rm h}(a)}_{\rm IT} \simeq {\bm L}_{\rm TT}({\bm q}(a),R(a))$. The stellar spin (Panel b) is consistent with the Lagrangian spins in Panel (e). This agreement suggests that the stellar component retains memory of the spin formed at an early epoch, which is well approximated by the tidal field on relatively small Lagrangian scales. In contrast, the gas spin (Panel c) is consistent with Lagrangian spins illustrated in Panel (f), indicating that the gas spin is more closely associated with the tidal field on larger Lagrangian scales. The large-scale tidal field in Lagrangian space naturally corresponds to material located in the outer protohalo region, which later supplies newly inflowing gas in Eulerian space through either mergers or smooth accretion. Thus, although the gas may enter the galaxy through different late-time physical channels, its spin is inherently encoded in the large-scale Lagrangian tidal field of the protohalo.

To quantify the evolution of the multi-component Eulerian spins (Panel g) and the various Lagrangian spins (Panel h), we plot the angular correlations (cosine) between the considered spin direction and a reference direction. The reference direction is chosen as ${\bm J}^{M_{\rm h}(a=1)}_{\rm L}$, meanwhile $\mu=\mp 1$ roughly corresponds to the final star/gas spin directions of the counter rotating system). 
For a specific component (gas or stars), ${\bm J}^{M(a)}$ is calculated using the respective particles within the main progenitor at epoch $a$.
When $a$ is small, the collapsed progenitor may contain too few particles for its spin direction to be reliably determined. We therefore plot $\mu( {\bm J}^{M(a)}, {\bm J}^{M_{\rm h}(a=1)}_{\rm L} )$ only when $M(a)$ contains at least $1000$ particles. The dot sizes indicate the corresponding spin parameter (described in Eqs. (\ref{eq.LambdaK}) - (\ref{eq.betaTT})). In Panel (g), before $a\simeq 0.71$, DM, gas and stellar components co-rotate, opposite to the reference direction. After that epoch, the newly accreted DM/gas reverses its direction, whereas stars maintain the memory of the earlier spin. In Panel (h), as viewed from a Lagrangian perspective, for all ${\bm J}^{M_{\rm h}(a)}_{\rm L}$, ${\bm J}^{M_{\rm h}(a)}_{\rm IT}$ and ${\bm L}_{\rm TT}({\bm q}(a),R(a))$ as a function of $a$, when $a\lesssim0.71$, the small scale tidal torque points toward the spin direction of stars. After this epoch, larger scale tidal torque reverses, pointing toward the final spins of DM and gas.

Other CRGs exhibit similar trends, which we do not detail here. For CRGs, we find that the characteristic spatial extents of the Lagrangian regions occupied by gas $R_{\rm large}$ is systematically larger than that of stars $R_{\rm small}$. Statistically, across the entire CRG sample, the gas spin is more positively correlated with the large-scale primordial spin field, with a mean correlation of $\langle \mu_{{\rm gas},R_{\rm large}}\rangle \simeq 0.55$ compared to $\langle \mu_{{\rm gas},R_{\rm small}}\rangle \simeq -0.02$ on small scales, and the stellar spin shows a stronger correlation with the small-scale spin field ($\langle \mu_{{\rm star},R_{\rm small}}\rangle \simeq 0.25$) than with the large-scale field ($\langle \mu_{{\rm star},R_{\rm large}}\rangle \simeq -0.30$). These results suggest that this scale-dependent correspondence unveils a universal origin of gas-stellar counter-rotation.


To distinguish the physical origin of counter-rotation from that of normal galaxies (subhaloID {\tt 480712}), we compare the above properties with a representative co-rotating galaxy in Figure 3. Unlike {\tt 393336}, the stellar and gaseous velocity fields exhibit consistent rotating directions. More importantly, the spin evolution differs from that shown in Figure 2. Throughout cosmic time, the DM, gas, and stellar components evolve coherently and have a common spin. There is no abrupt spin flip of the gas component. The Lagrangian spins remain closely aligned from early times to $a=1$, indicating that protohalo was embedded in a spatially coherent tidal field, where the large-scale environment exerted torques in the same direction as the small-scale.

\section{discussion and conclusions}
\label{sec:conclusions}

We investigate the origin of gas-stellar counter-rotation. The main findings are: (1) In the majority of CRGs, the stellar spin is anti-aligned with both the gas and DM spins. (2) The stellar component carries an earlier memory of galactic kinematics, acting as a fossil of the small-scale primordial tidal field, whereas the gas component reflects the large-scale primordial perturbations in Lagrangian space. 

Our results provide a further physical interpretation of these observational findings by tracing the origin of component-resolved angular momentum back to the primordial tidal field. While previous observational studies have established connections between gas-stellar misalignment, external gas accretion, and the surrounding cosmic environment, our results suggest that the different spins of gas and stars may already be imprinted in the initial conditions through the scale dependence of primordial tidal torques. This provides a physical connection between the observed kinematic misalignments of galaxies and cosmological initial conditions. Future observational validation depends on the accuracy of initial condition reconstruction, particularly on small scales, where uncertainties in the reconstructed tidal field may propagate into uncertainties in the predicted spin directions, while gas-stellar misalignment can in return serve as an additional observational constraint on the reconstruction itself.

\begin{acknowledgments}

We acknowledge the support from the National Natural Science Foundation of China grants Nos. 12673010, 12673013, 12333002, 12595311, 12133006; the China Manned Space Project CMS-CSST-2025-A08; the Fundamental Research Funds for the Central Universities Nos. 20720260041, KG202502; the National Key R\&D Program of China No. 2023YFA1607801; and the 111 Project No. B20019.

\end{acknowledgments}

\bibliography{references}{}

\begin{thebibliography}{}
\expandafter\ifx\csname natexlab\endcsname\relax\def\natexlab#1{#1}\fi
\providecommand{\url}[1]{\href{#1}{#1}}
\providecommand{\dodoi}[1]{doi:~\href{http://doi.org/#1}{\nolinkurl{#1}}}
\providecommand{\doeprint}[1]{\href{http://ascl.net/#1}{\nolinkurl{http://ascl.net/#1}}}
\providecommand{\doarXiv}[1]{\href{https://arxiv.org/abs/#1}{\nolinkurl{https://arxiv.org/abs/#1}}}

\bibitem[{M. {Bao} {et~al.}(2025){Bao}, {Chen}, {Gu}, {Wang}, {Shi}, \& {Wang}}]{Bao2025ApJL}
{Bao}, M., {Chen}, Y., {Gu}, Q., {et~al.} 2025, \bibinfo{title}{{The Large-scale Structure Supplies the Formation of Gas-star Misaligned Galaxies},} \apjl, 982, L29, \dodoi{10.3847/2041-8213/adbc68}

\bibitem[{M. {Beom} {et~al.}(2024){Beom}, {Walterbos}, \& {Bizyaev}}]{Beom2024AJ}
{Beom}, M., {Walterbos}, R. A.~M., \& {Bizyaev}, D. 2024, \bibinfo{title}{{SDSS. IV. MaNGA: The Impact of the Acquisition of Gas with Opposite Angular Momentum on the Evolution of Galaxies},} \aj, 168, 197, \dodoi{10.3847/1538-3881/ad6f0b}

\bibitem[{M.~C. {Bugue{\~n}o} {et~al.}(2026){Bugue{\~n}o}, {G{\'o}mez}, {Dolfi}, \& {Tissera}}]{Bugue2026A&A}
{Bugue{\~n}o}, M.~C., {G{\'o}mez}, F.~A., {Dolfi}, A., \& {Tissera}, P.~B. 2026, \bibinfo{title}{{Origins of coplanar counterrotating stellar disk components in late-type galaxies: I. Identification, characterization, and formation mechanisms of coplanar counterrotating stellar disk components},} \aap, 707, A51, \dodoi{10.1051/0004-6361/202557544}

\bibitem[{C. {Casanueva-Villarreal} {et~al.}(2026){Casanueva-Villarreal}, {Tissera}, {Padilla}, {Bignone}, {Gonzalez-Jara}, {Tapia-Contreras}, \& {Pedrosa}}]{C-V2026arXiv260607160C}
{Casanueva-Villarreal}, C., {Tissera}, P.~B., {Padilla}, N., {et~al.} 2026, \bibinfo{title}{{Angular momentum evolution in the CIELO simulations. I. Temporal evolution of gas-stellar misalignments and baryonic merger timing},} arXiv e-prints, arXiv:2606.07160, \dodoi{10.48550/arXiv.2606.07160}

\bibitem[{Y.-M. {Chen} {et~al.}(2016){Chen}, {Shi}, {Tremonti}, {Bershady}, {Merrifield}, {Emsellem}, {Jin}, {Huang}, {Fu}, {Wake}, {Bundy}, {Stark}, {Lin}, {Argudo-Fernandez}, {Bergmann}, {Bizyaev}, {Brownstein}, {Bureau}, {Chisholm}, {Drory}, {Guo}, {Hao}, {Hu}, {Li}, {Li}, {Roman Lopes}, {Pan}, {Riffel}, {Thomas}, {Wang}, {Westfall}, \& {Yan}}]{Chen2016NatCo}
{Chen}, Y.-M., {Shi}, Y., {Tremonti}, C.~A., {et~al.} 2016, \bibinfo{title}{{The growth of the central region by acquisition of counterrotating gas in star-forming galaxies},} Nature Communications, 7, 13269, \dodoi{10.1038/ncomms13269}

\bibitem[{A.~G. {Doroshkevich}(1970){Doroshkevich}}]{Doroshkevich1970Afz}
{Doroshkevich}, A.~G. 1970, \bibinfo{title}{{The space structure of perturbations and the origin of rotation of galaxies in the theory of fluctuation.},} Astrofizika, 6, 581

\bibitem[{D. {Gasymov} {et~al.}(2025){Gasymov}, {Katkov}, {Rubtsov}, {Saburova}, {Kniazev}, {Gelfand}, {Sil'chenko}, {Chilingarian}, {Moiseev}, {Kasparova}, \& {Zasov}}]{Gasymov2025ApJS}
{Gasymov}, D., {Katkov}, I.~Y., {Rubtsov}, E.~V., {et~al.} 2025, \bibinfo{title}{{Stellar CoRGI in MaNGA: Stellar Counterrotating Galaxies Identified in the MaNGA Survey},} \apjs, 281, 19, \dodoi{10.3847/1538-4365/ae0590}

\bibitem[{Y. {Jin} {et~al.}(2016){Jin}, {Chen}, {Shi}, {Tremonti}, {Bershady}, {Merrifield}, {Emsellem}, {Fu}, {Wake}, {Bundy}, {Lin}, {Argudo-Fernandez}, {Huang}, {Stark}, {Storchi-Bergmann}, {Bizyaev}, {Brownstein}, {Chisholm}, {Guo}, {Hao}, {Hu}, {Li}, {Li}, {Masters}, {Malanushenko}, {Pan}, {Riffel}, {Roman-Lopes}, {Simmons}, {Thomas}, {Wang}, {Westfall}, \& {Yan}}]{Jin2016MNRAS}
{Jin}, Y., {Chen}, Y., {Shi}, Y., {et~al.} 2016, \bibinfo{title}{{SDSS-IV MaNGA: properties of galaxies with kinematically decoupled stellar and gaseous components},} \mnras, 463, 913, \dodoi{10.1093/mnras/stw2055}

\bibitem[{S. {Khoperskov} {et~al.}(2021){Khoperskov}, {Zinchenko}, {Avramov}, {Khrapov}, {Berczik}, {Saburova}, {Ishchenko}, {Khoperskov}, {Pulsoni}, {Venichenko}, {Bizyaev}, \& {Moiseev}}]{Khoperskov2021MNRAS}
{Khoperskov}, S., {Zinchenko}, I., {Avramov}, B., {et~al.} 2021, \bibinfo{title}{{Extreme kinematic misalignment in IllustrisTNG galaxies: the origin, structure, and internal dynamics of galaxies with a large-scale counterrotation},} \mnras, 500, 3870, \dodoi{10.1093/mnras/staa3330}

\bibitem[{P. {Motloch} {et~al.}(2021){Motloch}, {Yu}, {Pen}, \& {Xie}}]{Motloch2021NA}
{Motloch}, P., {Yu}, H.-R., {Pen}, U.-L., \& {Xie}, Y. 2021, \bibinfo{title}{{An observed correlation between galaxy spins and initial conditions},} Nature Astronomy, 5, 283, \dodoi{10.1038/s41550-020-01262-3}

\bibitem[{D. {Nelson} {et~al.}(2019){Nelson}, {Springel}, {Pillepich}, {Rodriguez-Gomez}, {Torrey}, {Genel}, {Vogelsberger}, {Pakmor}, {Marinacci}, {Weinberger}, {Kelley}, {Lovell}, {Diemer}, \& {Hernquist}}]{Nelson2019ComAC}
{Nelson}, D., {Springel}, V., {Pillepich}, A., {et~al.} 2019, \bibinfo{title}{{The IllustrisTNG simulations: public data release},} Computational Astrophysics and Cosmology, 6, 2, \dodoi{10.1186/s40668-019-0028-x}

\bibitem[{O. {Osman} {\&} K. {Bekki}(2017){Osman} \& {Bekki}}]{Osman2017MNRAS}
{Osman}, O., \& {Bekki}, K. 2017, \bibinfo{title}{{Strong suppression of star formation and spiral arm formation in disc galaxies with counter-rotating gas discs},} \mnras, 471, L87, \dodoi{10.1093/mnrasl/slx104}

\bibitem[{P.~J.~E. {Peebles}(1969){Peebles}}]{Peebles1969ApJ}
{Peebles}, P.~J.~E. 1969, \bibinfo{title}{{Origin of the Angular Momentum of Galaxies},} \apj, 155, 393, \dodoi{10.1086/149876}

\bibitem[{A. {Pillepich} {et~al.}(2018){Pillepich}, {Nelson}, {Hernquist}, {Springel}, {Pakmor}, {Torrey}, {Weinberger}, {Genel}, {Naiman}, {Marinacci}, \& {Vogelsberger}}]{Pillepich2018MNRAS}
{Pillepich}, A., {Nelson}, D., {Hernquist}, L., {et~al.} 2018, \bibinfo{title}{{First results from the IllustrisTNG simulations: the stellar mass content of groups and clusters of galaxies},} \mnras, 475, 648, \dodoi{10.1093/mnras/stx3112}

\bibitem[{C. {Porciani} {et~al.}(2002){Porciani}, {Dekel}, \& {Hoffman}}]{Porciani2002MNRAS}
{Porciani}, C., {Dekel}, A., \& {Hoffman}, Y. 2002, \bibinfo{title}{{Testing tidal-torque theory - I. Spin amplitude and direction},} \mnras, 332, 325, \dodoi{10.1046/j.1365-8711.2002.05305.x}

\bibitem[{M.-J. {Sheng} {et~al.}(2024){Sheng}, {Zhu}, {Yu}, {Ma}, {Li}, {Wang}, \& {Kang}}]{Sheng2024PhRvD}
{Sheng}, M.-J., {Zhu}, L., {Yu}, H.-R., {et~al.} 2024, \bibinfo{title}{{Spin speed correlations and the evolution of galaxy-halo systems},} \prd, 109, 123548, \dodoi{10.1103/PhysRevD.109.123548}

\bibitem[{M.-J. {Sheng} {et~al.}(2023){Sheng}, {Yu}, {Li}, {Liao}, {Du}, {Wang}, {Wang}, {Xu}, {Genel}, \& {Irodotou}}]{Sheng2023ApJ}
{Sheng}, M.-J., {Yu}, H.-R., {Li}, S., {et~al.} 2023, \bibinfo{title}{{Baryonic Effects on Lagrangian Clustering and Angular Momentum Reconstruction},} \apj, 943, 128, \dodoi{10.3847/1538-4357/acae92}

\bibitem[{M.-J. {Sheng} {et~al.}(2026){Sheng}, {Yu}, {Bao}, {Chen}, {Shao}, {Guo}, {Chen}, {Wang}, {Pen}, {Wang}, \& {Yang}}]{2026ShengNA}
{Sheng}, M.-J., {Yu}, H.-R., {Bao}, M., {et~al.} 2026, \bibinfo{title}{{A high-significance detection of primordial tidal torque imprints},} Nature Astronomy, \dodoi{10.1038/s41550-026-02948-w}

\bibitem[{V. {Springel} {et~al.}(2018){Springel}, {Pakmor}, {Pillepich}, {Weinberger}, {Nelson}, {Hernquist}, {Vogelsberger}, {Genel}, {Torrey}, {Marinacci}, \& {Naiman}}]{Springel2018MNRAS}
{Springel}, V., {Pakmor}, R., {Pillepich}, A., {et~al.} 2018, \bibinfo{title}{{First results from the IllustrisTNG simulations: matter and galaxy clustering},} \mnras, 475, 676, \dodoi{10.1093/mnras/stx3304}

\bibitem[{T.~K. {Starkenburg} {et~al.}(2019){Starkenburg}, {Sales}, {Genel}, {Manzano-King}, {Canalizo}, \& {Hernquist}}]{Starkenburg2019ApJ}
{Starkenburg}, T.~K., {Sales}, L.~V., {Genel}, S., {et~al.} 2019, \bibinfo{title}{{On the Origin of Star-Gas Counterrotation in Low-mass Galaxies},} \apj, 878, 143, \dodoi{10.3847/1538-4357/ab2128}

\bibitem[{H. {Wang} {et~al.}(2014){Wang}, {Mo}, {Yang}, {Jing}, \& {Lin}}]{Wang2014ApJ}
{Wang}, H., {Mo}, H.~J., {Yang}, X., {Jing}, Y.~P., \& {Lin}, W.~P. 2014, \bibinfo{title}{{ELUCID{\textemdash}Exploring the Local Universe with the Reconstructed Initial Density Field. I. Hamiltonian Markov Chain Monte Carlo Method with Particle Mesh Dynamics},} \apj, 794, 94, \dodoi{10.1088/0004-637X/794/1/94}

\bibitem[{H. {Wang} {et~al.}(2016){Wang}, {Mo}, {Yang}, {Zhang}, {Shi}, {Jing}, {Liu}, {Li}, {Kang}, \& {Gao}}]{2016WangApJ}
{Wang}, H., {Mo}, H.~J., {Yang}, X., {et~al.} 2016, \bibinfo{title}{{ELUCID - Exploring the Local Universe with ReConstructed Initial Density Field III: Constrained Simulation in the SDSS Volume},} \apj, 831, 164, \dodoi{10.3847/0004-637X/831/2/164}

\bibitem[{S.~D.~M. {White}(1984){White}}]{White1984ApJ}
{White}, S.~D.~M. 1984, \bibinfo{title}{{Angular momentum growth in protogalaxies},} \apj, 286, 38, \dodoi{10.1086/162573}

\bibitem[{Q. {Wu} {et~al.}(2021){Wu}, {Yu}, {Liao}, \& {Du}}]{Wu2021PhRvD}
{Wu}, Q., {Yu}, H.-R., {Liao}, S., \& {Du}, M. 2021, \bibinfo{title}{{Spin mode reconstruction in Lagrangian space},} \prd, 103, 063522, \dodoi{10.1103/PhysRevD.103.063522}

\bibitem[{H. {Xu} {et~al.}(2022){Xu}, {Chen}, {Shi}, {Zhou}, {Bizyaev}, {Bao}, {Beom}, {Fern{\'a}ndez-Trincado}, \& {Cao}}]{Xu2022MNRAS}
{Xu}, H., {Chen}, Y., {Shi}, Y., {et~al.} 2022, \bibinfo{title}{{SDSS-IV MaNGA: spatial resolved properties of kinematically misaligned galaxies},} \mnras, 511, 4685, \dodoi{10.1093/mnras/stac354}

\bibitem[{D.~G. {York} {et~al.}(2000){York}, {Adelman}, {Anderson}, {Anderson}, {Annis}, {Bahcall}, {Bakken}, {Barkhouser}, {Bastian}, {Berman}, {Boroski}, {Bracker}, {Briegel}, {Briggs}, {Brinkmann}, {Brunner}, {Burles}, {Carey}, {Carr}, {Castander}, {Chen}, {Colestock}, {Connolly}, {Crocker}, {Csabai}, {Czarapata}, {Davis}, {Doi}, {Dombeck}, {Eisenstein}, {Ellman}, {Elms}, {Evans}, {Fan}, {Federwitz}, {Fiscelli}, {Friedman}, {Frieman}, {Fukugita}, {Gillespie}, {Gunn}, {Gurbani}, {de Haas}, {Haldeman}, {Harris}, {Hayes}, {Heckman}, {Hennessy}, {Hindsley}, {Holm}, {Holmgren}, {Huang}, {Hull}, {Husby}, {Ichikawa}, {Ichikawa}, {Ivezi{\'c}}, {Kent}, {Kim}, {Kinney}, {Klaene}, {Kleinman}, {Kleinman}, {Knapp}, {Korienek}, {Kron}, {Kunszt}, {Lamb}, {Lee}, {Leger}, {Limmongkol}, {Lindenmeyer}, {Long}, {Loomis}, {Loveday}, {Lucinio}, {Lupton}, {MacKinnon}, {Mannery}, {Mantsch}, {Margon}, {McGehee}, {McKay}, {Meiksin}, {Merelli}, {Monet}, {Munn}, {Narayanan}, {Nash}, {Neilsen}, {Neswold}, {Newberg}, {Nichol},
  {Nicinski}, {Nonino}, {Okada}, {Okamura}, {Ostriker}, {Owen}, {Pauls}, {Peoples}, {Peterson}, {Petravick}, {Pier}, {Pope}, {Pordes}, {Prosapio}, {Rechenmacher}, {Quinn}, {Richards}, {Richmond}, {Rivetta}, {Rockosi}, {Ruthmansdorfer}, {Sandford}, {Schlegel}, {Schneider}, {Sekiguchi}, {Sergey}, {Shimasaku}, {Siegmund}, {Smee}, {Smith}, {Snedden}, {Stone}, {Stoughton}, {Strauss}, {Stubbs}, {SubbaRao}, {Szalay}, {Szapudi}, {Szokoly}, {Thakar}, {Tremonti}, {Tucker}, {Uomoto}, {Vanden Berk}, {Vogeley}, {Waddell}, {Wang}, {Watanabe}, {Weinberg}, {Yanny}, {Yasuda}, \& {SDSS Collaboration}}]{York2000AJ}
{York}, D.~G., {Adelman}, J., {Anderson}, Jr., J.~E., {et~al.} 2000, \bibinfo{title}{{The Sloan Digital Sky Survey: Technical Summary},} \aj, 120, 1579, \dodoi{10.1086/301513}

\bibitem[{H.-R. {Yu} {et~al.}(2020){Yu}, {Motloch}, {Pen}, {Yu}, {Wang}, {Mo}, {Yang}, \& {Jing}}]{Yu2020PhRvL}
{Yu}, H.-R., {Motloch}, P., {Pen}, U.-L., {et~al.} 2020, \bibinfo{title}{{Probing Primordial Chirality with Galaxy Spins},} \prl, 124, 101302, \dodoi{10.1103/PhysRevLett.124.101302}

\bibitem[{Q. {Zhang} {et~al.}(2025){Zhang}, {Bao}, {Chen}, {Yu}, {Shi}, \& {Ho}}]{zhang2025ApJ}
{Zhang}, Q., {Bao}, M., {Chen}, Y., {et~al.} 2025, \bibinfo{title}{{The Impact of External Gas Accretion on the Distribution of H I Gas in Galaxies},} \apj, 993, 149, \dodoi{10.3847/1538-4357/ae07d6}

\bibitem[{Y. {Zhou} {et~al.}(2022{\natexlab{a}}){Zhou}, {Chen}, {Shi}, {Bizyaev}, {Guo}, {Bao}, {Xu}, {Yu}, \& {Brownstein}}]{Zhou2022MNRAS}
{Zhou}, Y., {Chen}, Y., {Shi}, Y., {et~al.} 2022{\natexlab{a}}, \bibinfo{title}{{SDSS-IV MaNGA: global properties of kinematically misaligned galaxies},} \mnras, 515, 5081, \dodoi{10.1093/mnras/stac2016}

\bibitem[{I.~A. {Zinchenko}(2023){Zinchenko}}]{Zinchenko2023A&A}
{Zinchenko}, I.~A. 2023, \bibinfo{title}{{Gas and stellar kinematic misalignment in MaNGA galaxies: What is the origin of counter-rotating gas?},} \aap, 674, L7, \dodoi{10.1051/0004-6361/202346846}

\end{thebibliography}
\bibliographystyle{aasjournalv7.1}

\end{document}